\documentclass[sn-mathphys-num]{sn-jnl}

\usepackage[T1]{fontenc}
\usepackage[latin9]{inputenc}
\usepackage{amsmath}
\usepackage{amsthm}
\usepackage{amssymb}
\usepackage{cancel}

\usepackage{bm}
\usepackage{subcaption}

\def\br{\bm{\rho}}

\begin{document}

\title[Robustness of chaotic-light correlation imaging against turbulence]{Robustness of chaotic-light correlation imaging against turbulence}

\author[1,2]{\fnm{Giovanni} \sur{Scala}}
\author[2,3]{\fnm{Gianlorenzo} \sur{Massaro}}
\author[3]{\fnm{Germano} \sur{Borreggine}}
\author[1,2]{\fnm{Cosmo} \sur{Lupo}}
\author[2,3]{\fnm{Milena} \sur{D'Angelo}}
\author[2,3]{\fnm{Francesco V.} \sur{Pepe}}

\affil[1]{\orgname{Dipartimento Interateneo di Fisica, Politecnico di Bari}, \orgaddress{\city{Bari}, \postcode{I-70125}, \country{Italy}}}
\affil[2]{\orgname{INFN - Sezione di Bari}, \orgaddress{\city{Bari}, \postcode{I-70125}, \country{Italy}}}
\affil[3]{\orgname{Dipartimento Interateneo di Fisica, Universit\`a degli Studi di Bari}, \orgaddress{\city{Bari}, \postcode{I-70125}, \country{Italy}}}
\affil[*]{Corresponding author: milena.dangelo@uniba.it}

\abstract{We consider an imaging scheme, inspired by microscopy, in which both correlation imaging and first-order intensity imaging can be performed simultaneously, to investigate the effects of strong turbulence on the two different kinds of images. The comparison between direct and correlation imaging in the presence of strong turbulence unambiguously revealed an advantage of the latter. Remarkably, this advantage, quantified by analyzing the visibility of periodic sample patterns, is more striking when the presence of turbulence becomes the dominant factor in determining the image resolution.}

\keywords{Turbulence, imaging, correlation imaging, chaotic light}

\maketitle

\section{Introduction}

The propagation of electromagnetic fields through various media is of considerable interest due to its wide-ranging applications in imaging and communication technologies \cite{gbur2002spreading,schulz2005optimal}. Many imaging systems, including terrestrial and satellite imaging, underwater imaging, and microscopy through biological tissues, are significantly impacted by the non-homogeneous nature of the propagation medium. The primary factors influencing light propagation in such media include absorption, scattering, and fluctuations in the refractive index due to optical turbulence \cite{Fante1985}. These turbulence-induced fluctuations lead to wavefront distortions, causing phenomena such as beam spreading, beam wander, and a loss of spatial coherence of the electromagnetic field. These distortions impose detrimental limitations on the angular resolution of optical systems. 

In recent years, the impact of turbulence on quantum optical technologies, particularly those utilizing entangled photons, has attracted significant attention. Technologies such as quantum key distribution \cite{Pirandola2020,Zapatero2023} or quantum sensing based on intensity correlations \cite{cassano2017spatial,dangelo2017characterization,pepe2022distance,massaro2022lightfield} exploit non-local properties of light and rely on measuring radiation impinging on spatially separated sensors. This study aims at evaluating the imaging performance of an intensity correlation imaging technique under strong atmospheric turbulence, a scenario frequently encountered in long-range imaging and free-space optical communication, in which the phase disturbance due to the turbulent medium tends to have zero average \cite{Fante1985}. Specifically, we consider a case of correlation plenoptic imaging (CPI), initially introduced as an enhancement of ghost imaging \cite{pittman1995optical,bennink2002two,valencia2005two,gatti2004ghost,scarcelli2006can,osullivan2010comparison,brida2011systematic,cassano2017spatial,dangelo2017characterization} capable of reconstructing not only a planar image, but rather a volumetric light distribution \cite{dangelo2016correlation,pepe2016correlation,pepe2017diffraction,cpi_jopt}, by reconstructing the direction of light propagating in a given scene. Much research has already been devoted to determining the robustness of correlation imaging methods against turbulence and scattering environments \cite{cheng2009ghost,li2010ghost,chan2011theoretical,hardy2011reflective,dixon2011quantum,meyers2011turbulence,shi2012adaptive,erkmen2012computational}. 
Unlike the first CPI proposal based on ghost imaging \cite{dangelo2016correlation}, imaging an absorptive target placed in one optical arm, our analysis is carried out in a CPI scheme in which the object lies in the common path of the correlated beams, and can thus be treated as a source of chaotic light \cite{massaro2022lightfield,massaro2023correlated,cpiap,scagliola2020correlation,abbattista2021towards}. Such a scheme can be applied not only to engineered laboratory sources, but also to natural sources that can be unavoidably surrounded by a turbulent media. While here we shall consider the case of strong turbulence, a previous analysis investigated the effect of quasi-static turbulence with arbitrary spatial features in a similar scheme \cite{pepe20243d}. 

The article is organized as follows. In Section~\ref{sec:modeling}, we describe the optical scheme considered in the analysis and introduce the model of turbulence to compute the intensity correlation functions. Such a task is technically challenging, due to the involvement of four-point field correlations, which require specific assumptions of turbulence statistics, that are not required when disturbance is present in only one of the correlated optical paths. Under these assumptions, the general form of the correlation functions in the presence of turbulence are derived. In Section~\ref{sec:quality}, we compare the quality of CPI images and standard images, based on direct intensity measures, using as a test object a planar sample emitting chaotic light with a periodic intensity profile. In Section~\ref{sec:conclusions}, we discuss the relevance of our findings, pointing towards a better robustness of correlation imaging against turbulence, and provide an outlook to generalization and extensions of the present analysis.

\section{Strong turbulence modeling in correlation imaging}\label{sec:modeling}

\begin{figure}
    \centering
    \includegraphics[width=\textwidth]{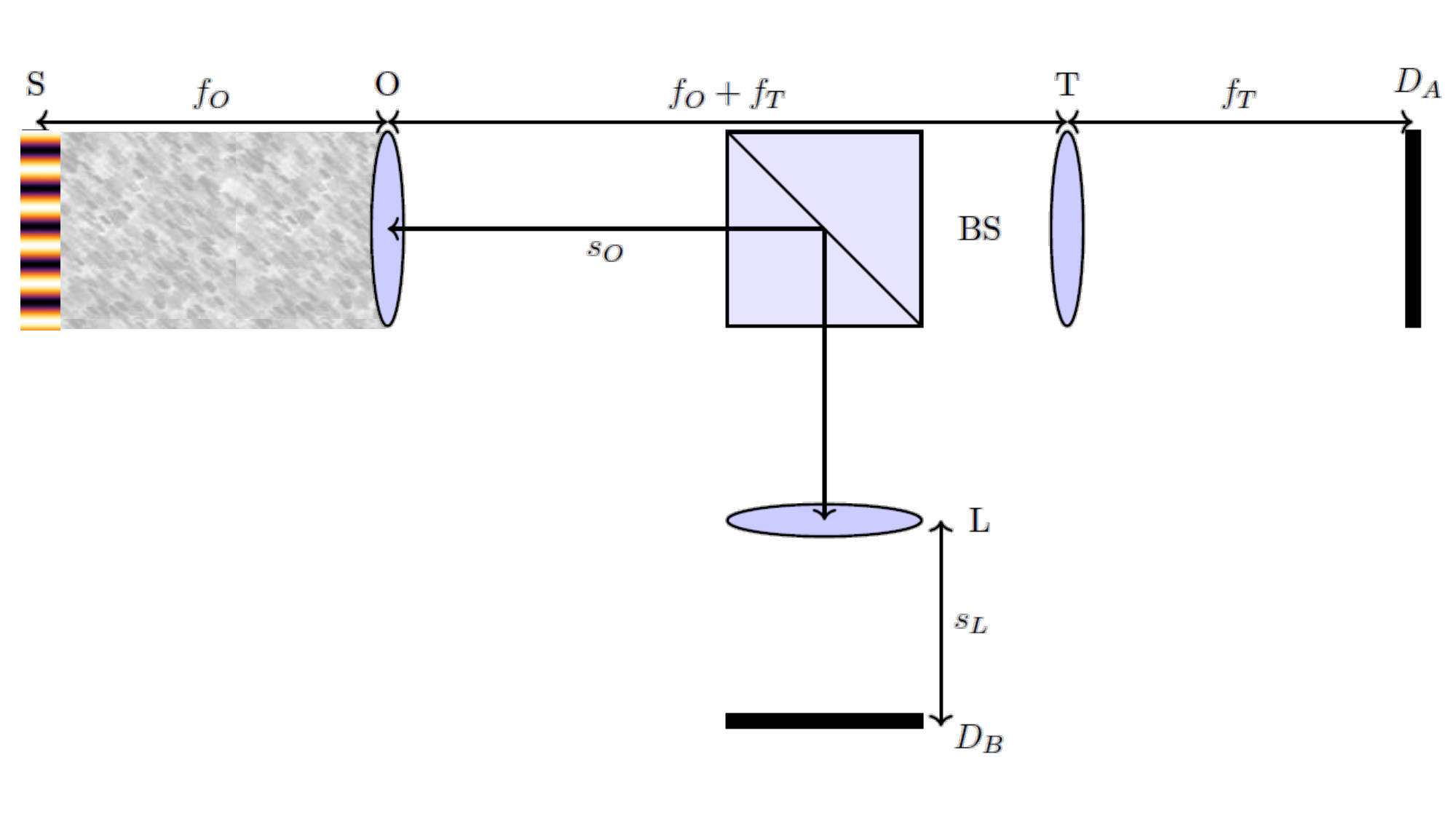}
    \caption{Optical scheme considered in the derivation of direct and correlation imaging properties. Light emitted from a planar sample S impinges on an objective lens O after propagating through a turbulent volume. After O, light is split in two paths by a beam splitter: along the transmitted path ($a$), a tube lens T focuses an image of the sample on the detector $\mathrm{D}_a$, as in an ordinary microscope; along the reflected path ($b$), an additional lens L focuses an image of O on the sensor $\mathrm{D}_b$.}
    \label{fig:opticalscheme}
\end{figure}

Throughout our analysis, we shall refer to the correlation plenoptic microscopy scheme, described in Ref.~\cite{scagliola2020correlation} and experimentally realized in Ref.~\cite{massaro2022lightfield}. Light emitted from a planar sample S, treated as a pseudothermal source of quasi-monochromatic light with wavenumber $k$, propagates through a turbulent volume before being collected by the objective lens O, placed at a focal distance $f_O$ from S. After O, light is divided into two distinct paths $A$ and $B$. Along $A$, as in a standard microscope operating in a $4f$-configuration, a tube lens T of focal length $f_T$ is placed at an equivalent optical distance $f_O+f_T$ from the objective, with a spatially resolving detector $\mathrm{D}_a$ downstream, at a distance $f_T$. In this way, an image of the sample is formed on the $\mathrm{D}_a$, with magnification $M_T=f_T/f_O$. Along path $B$, a lens L of focal length $f_L$ is placed in between O and another spatially resolving sensor $\mathrm{D}_b$, at optical distances $d_{o\ell}$ from the former and $d_{\ell}$ from the latter, in such a way that $1/d_{o\ell}+1/d_{\ell}=1/f_L$. In this way, an image of O with absolute magnification $M_L=d_{\ell}/d_{o\ell}$ is reproduced on $\mathrm{D}_b$.

In each frame, the detectors $\mathrm{D}_a$ and $\mathrm{D}_b$ acquire samples of the intensity patterns $\mathcal{I}_A(\br_a)$ and $\mathcal{I}_A(\br_b)$, respectively, where $\br_{a,b}$ are the pixel coordinates on the given detector planes. Assuming ergodicity, the signals collected by each pixel and averaged in time converge to the mean intensity $I_A(\br_a) = \langle \mathcal{I}_A(\br_a)\rangle$ and $I_B(\br_b) = \langle \mathcal{I}_B(\br_b)\rangle$, with the angular brackets denoting an average on light statistics. A plenoptic image of the scene is encoded in the correlations between intensity fluctuations \cite{massaro2022effect}
\begin{equation}\label{eq:Gamma}
\Gamma (\br_a, \br_b) = \langle \Delta \mathcal{I}_A(\br_a) \Delta \mathcal{I}_B(\br_b) \rangle = \langle \mathcal{I}_A(\br_a) \mathcal{I}_B(\br_b) \rangle - I_A(\br_a) I_B(\br_b) 
\end{equation}
We adopt a scalar approximation of the electromagnetic field, based on the assumption that polarization is not relevant for the described phenomenology. Therefore, intensities will be derived as the square modulus of scalar field components $V_A(\br_a)$ and $V_B(\br_b)$. While randomness usually comes only from the source statistics, here the angular brackets must also involve the operation of averaging on the \textit{turbulence} statistics, which requires a model to be appropriately described.

Turbulence plays its crucial role in how the field on the source, characterized by the spatial distribution $V_S(\br_s)$, propagates towards the objective lens. Under the aforementioned assumptions, this value can be expressed as 
\begin{equation}
    V_O(\br_o) = \frac{-i k}{2\pi f_O} \int d^2\br_s \exp\left( \frac{i k}{2 f_O} (\br_o - \br_s)^2 + \psi(\br_o,\br_s) \right) V_S(\br_s) ,
\end{equation}
where we have arbitrarily factorized the free propagator, depending only on the distance between $\br_o$ and $\br_s$, leaving all the corrections due to turbulence in the complex quantity $\psi(\br_s,\br_o)$, that is generally called the \textit{Rytov phase} \cite{Fante1985} and accounts for both absorption (real part) and dispersion (imaginary part). In this article, we focus on the case of \textit{strong turbulence}, where we assume $C_n^2 k^{7/6} f_O^{11/6}\gg 1$, with $C_n$ an $O(1)$ structure constant \cite{Fante1985}. In this conditions, the quantity $e^{\psi(\br_o,\br_s)}$ can be modelled as a zero-average Gaussian random variable, approximately characterized by the two-phase correlation \cite{Fante1985}
\begin{equation}\label{eq:turb2p}
    \left\langle e^{ \psi(\br_o,\br_s) + \psi^*(\br'_o,\br'_s) }  \right\rangle = C_T \exp \left( - \frac{ (\br_s - \br'_s)^2 + (\br_o - \br'_o)^2 + (\br_s - \br'_s)\cdot(\br_o - \br'_o) }{2\sigma_T^2}  \right) ,
\end{equation}
with expectation values of the type $\langle e^{ \psi(\br_o,\br_s) + \psi(\br'_o,\br'_s) } \rangle$ vanishing. The quantity $C_T$ is related to absorption, that will be considered independent of the initial and final points, while $\sigma_T$ represents the phase correlation length of turbulence, and will play the most relevant role in our analysis. After the objective, characterized by the effective pupil function $P_O(\br_o)$, the field propagates towards the detector $\mathrm{D}_a$ on one arm
\begin{equation}
    V_A(\br_a) = \int d^2\br_o \int d^2\br_t V_O(\br_o) P_O (\br_o) e^{ - \frac{i k}{2 f_O} \br_o^2 + \frac{i k}{2(f_O+f_t)} (\br_t-\br_o)^2 - \frac{i k}{2 f_t} \br_t^2 + \frac{i k}{2 f_t} (\br_a-\br_t)^2 }
\end{equation}
where we neglect the finite size of the tube lens, and towards $\mathrm{D}_b$ on the other arm
\begin{equation}
    V_B(\br_b) = \int d^2\br_o \int d^2\br_{\ell} V_O(\br_o) P_O(\br_O) e^{ - \frac{i k}{2 f_O} \br_o^2 + \frac{i k}{2 d_{o\ell}} (\br_{\ell}-\br_o)^2 - \frac{i k}{2 f_L} \br_{\ell}^2 + \frac{i k}{2 d_{\ell}} (\br_b-\br_{\ell})^2 } , 
\end{equation}
where the finite extent of the imaging lens $L$ is neglected. Due to the focusing condition $1/d_{o\ell} + 1/d_{\ell} = 1/f_L$, the field on the detector $\mathrm{D}_b$ is proportional to $V_O(-\br_b/M_L)$, with $M_L = d_{\ell}/d_{o\ell}$ the absolute magnification provided by the lens $L$.

A computation of the two-point intensity correlation function \eqref{eq:Gamma} involves terms of the kind
\begin{equation}\label{eq:4point}
    \left\langle e^{ \psi(\br_o,\br_s) + \psi^*(\br'_o,\br'_s) + \psi(\br''_o,\br''_s) + \psi^*(\br'''_o,\br'''_s) }  \right\rangle 
    \left\langle V_S(\br_s) V_S^*(\br'_s) V_S(\br''_s) V_S^*(\br'''_s) \right\rangle ,
\end{equation}
where factorization occurs due to the independence of fluctuations determined by to the source statistics and by turbulence. The evaluation of the above quantity is simplified by assuming that both statistics are Gaussian and by neglecting the coherence area on the source plane:
\begin{equation}
    \langle V_S(\br_s) V_S(\br'_s) \rangle \sim \delta(\br_s-\br'_s) I_S(\br_s) ,
\end{equation}
where $I_S=\langle |V_S|^2 \rangle$ is the source intensity profile. In this situation, only the terms
\begin{multline}
    I_S(\br_s) I_S (\br''_s) \left( \left\langle e^{ \psi(\br_o,\br_s) + \psi^*(\br'_o,\br_s)} \right\rangle \left\langle e^{ \psi(\br''_o,\br''_s) + \psi^*(\br'''_o,\br''_s) } \right\rangle \right. \\
    + \left. \left\langle e^{ \psi(\br_o,\br_s) + \psi^*(\br'''_o,\br_s)} \right\rangle \left\langle e^{ \psi(\br''_o,\br''_s) + \psi^*(\br'_o,\br''_s) } \right\rangle \right)
\end{multline}
give a relevant contribution to the four-point function, since turbulence correlators of the kind
\begin{equation}
    \left\langle e^{ \psi(\br_o,\br_s) + \psi^*(\br'_o,\br_s)} \right\rangle = C_T \exp\left( - \frac{(\br_o - \br'_o)^2}{2\sigma_T^2} \right) 
\end{equation}
do not limit the integration range in the $(\br_s,\br''_s)$ space, while the remaining contributions
\begin{multline}
    I_S(\br_s) I_S (\br''_s) \left( \left\langle e^{ \psi(\br_o,\br_s) + \psi^*(\br'''_o,\br''_s)} \right\rangle \left\langle e^{ \psi(\br''_o,\br''_s) + \psi^*(\br'_o,\br_s) } \right\rangle \right. \\
    + \left. \left\langle e^{ \psi(\br_o,\br_s) + \psi^*(\br'_o,\br''_s)} \right\rangle \left\langle e^{ \psi(\br''_o,\br''_s) + \psi^*(\br'''_o,\br_s) } \right\rangle \right)
\end{multline}
to \eqref{eq:4point} are suppressed like $\sigma_T^2$ over the area of the object. Under the discussed assumptions, the intensity correlation \eqref{eq:Gamma} takes the simple form
\begin{equation}
    \Gamma (\br_a,\br_b) = \left| \left\langle V_A(\br_a) V_B^*(\br_b) \right\rangle \right|^2 ,
\end{equation}
which interestingly allows to recover, in the strong turbulence regime, the same form that holds in the absence of turbulence.

Based on the above considerations, the evaluation of the correlation function for arbitrary sample intensity profiles and objective pupils leads to
\begin{equation}\label{eq:Gammaapprox}
    \Gamma(\br_a,\br_b) = \left| P_O \left( - \frac{\br_b}{M_L} \right) \int d^2\br_s \int d^2\br_o I_S(\br_s) P_O(\br_o) \exp\left[ \Psi(\br_a,\br_b,\br_s,\br_o) \right] \right|^2 ,
\end{equation}
with
\begin{equation}
\Psi(\br_a,\br_b,\br_s,\br_o)  = - \frac{1}{2\sigma_T^{2}} \left( \br_O + \frac{\br_b}{M_L} \right)^2 - i k \left[ \left( \br_s + \frac{\br_a}{M_T} \right) \cdot \frac{\br_o}{f_O}  + \frac{\br_b\cdot\br_s}{f_{O}M_{L}} \right] .
\end{equation}
In the following, we will assume an effective Gaussian field transmission profile
\begin{equation}
    P_O(\br_o) = \exp \left( - \frac{\br_o^2}{2\sigma_O^2} \right)
\end{equation}
for the objective lens, which will enable to analytically evaluate image visibilities in our case study, thus providing physically insightful results.

\section{Image quality of periodic intensity patterns}\label{sec:quality}

To test the combined dependence of image quality on turbulence and on the natural resolution defined by the lens aperture, we consider periodic intensity patterns
\begin{equation}\label{eq:periodic}
    I_S(\br_s) = I_0 \cos^2 \left( \frac{x_s}{2 w_s} \right) ,
\end{equation}
modulated along the $x$ direction and characterized by a sequence of peaks at a distance $\pi w_s$ from each other. For a comparison between the length scales involved in the process, it is convenient to introduce the coherence length 
\begin{equation}
    \sigma_c = \frac{f_O}{k w_S}
\end{equation}
associated with the propagation by a distance $f_O$ (namely, from the source to the objective lens) of light emitted from incoherent sources of a size comparable with $w_s$. Considering the intensity profile \eqref{eq:periodic} and the form \eqref{eq:Gammaapprox}, derived under assumptions on turbulence, the intensity correlation function reads
\begin{multline}\label{eq:GammaRes}
    \Gamma \left(\br_a, \br_b\right) = \Gamma(0,0) \, \mathrm{e}^{-\frac{2 \br_{b}^{2}}{\sigma_{o}^{2}M_{L}^{2}}} \Biggl[ \left(1+\mathrm{e}^{-\frac{N_{T}^{2}+N_{O}^{2}}{2}}\cosh\left(\frac{N_{O}x_{b}}{M_{L}\sigma_{O}}\right)\cos\left(\frac{x_{a}}{M_{T}w_{S}}\right)\right)^{2} \\ + \mathrm{e}^{-\left(N_{T}^{2}+N_{O}^{2}\right)} \sinh^{2}\left(\frac{N_{O}x_{b}}{M_{L}\sigma_{o}}\right)\sin^{2}\left(\frac{x_{a}}{M_{T}w_{S}}\right)\Biggr] ,
\end{multline}
which is, as expected, independent of $y_a$, with $y_b$ appearing only in the Gaussian envelope. In the above expression, we introduced the dimensionless quantities
\begin{equation}
    N_{O}=\frac{f_{O}}{w_s k\sigma_{O}} = \frac{\sigma_c}{\sigma_O} , \quad
    N_{T}=\frac{f_{O}}{w_sk\sigma_{T}} = \frac{\sigma_c}{\sigma_T} ,
\end{equation}
that allow to highlight the relevant physical scales. Indeed, the results crucially depend on the comparison of the coherence length $\sigma_C$ with the lens size and with the turbulence correlation length, which sets the length scale on which the phase disturbance determined by turbulence is roughly homogeneous. It is worth noticing that the quantity $N_0$ can also be interpreted as the number of objective resolution cells (of size $f_O/k\sigma_O$) that fit into a distance $w_s$. Therefore, when $N_O\sim 1$, the peaks of the intensity profiles are expected to become unresolved.

In the context of CPI, by varying $\br_b$ one obtains a collection of images $\Gamma(\br_a,\br_b)$, that can be properly realigned and integrated to obtain a \textit{refocused} image of an off-focus object, exploiting the whole signal transmitted by the lens instead that the small lens area corresponding to the pixel around $\br_b$ \cite{massaro2022refocusing}. While not substantially changing the resolution of off-focus objects, such a procedure enables to improve the signal-to-noise ratio and provides an axial sectioning that is absent in the single $\Gamma(\br_a,\br_b)$ \cite{massaro2024assessing}. Our case study is simplified by the object being in focus, thus not requiring any realigning operation: all the redundant images obtained from different points on $\mathrm{D}_b$ can be integrated into
\begin{equation}
    \Sigma(\br_a) = \int d^2\br_b \Gamma( \br_a, \br_b ) .
\end{equation}
The result for the considered class of periodic objects reads
\begin{equation}\label{eq:SigmaRes}
    \Sigma(\br_a) = \Sigma(0) \left[1+2\mathrm{e}^{-\frac{3N_{O}^{2}+4N_{T}^{2}}{8}}\cos\left(\frac{x_a}{M_{T}w_{s}}\right)+\frac{\mathrm{e}^{-N_{T}^{2}-N_{O}^{2}}}{2}\left(\cos\left(\frac{2 x_a}{M_{T}w_{s}}\right)+\mathrm{e}^{\frac{N_{O}^{2}}{2}}\right)\right] .
\end{equation}
The expressions of the pointwise correlation \eqref{eq:GammaRes} and the integrated correlation \eqref{eq:SigmaRes} can be compared with the direct intensity measurement performed on $\mathrm{D}_a$, which yields
\begin{equation}\label{eq:IRes}
    I_{A} (\br_a) = \langle V_A^*(\br_a) V_A (\br_a) \rangle = I_A(0) \left[ 1+ \mathrm{e}^{-\frac{N_{O}^{2}+2 N_{T}^{2}}{4}} \cos \left( \frac{x_{a}}{w_{S}M_{T}} \right) \right] .
\end{equation}

\begin{figure}
    \centering
    \begin{subfigure}[h]{.49\textwidth}
     \centering 
     \includegraphics[width=\textwidth]{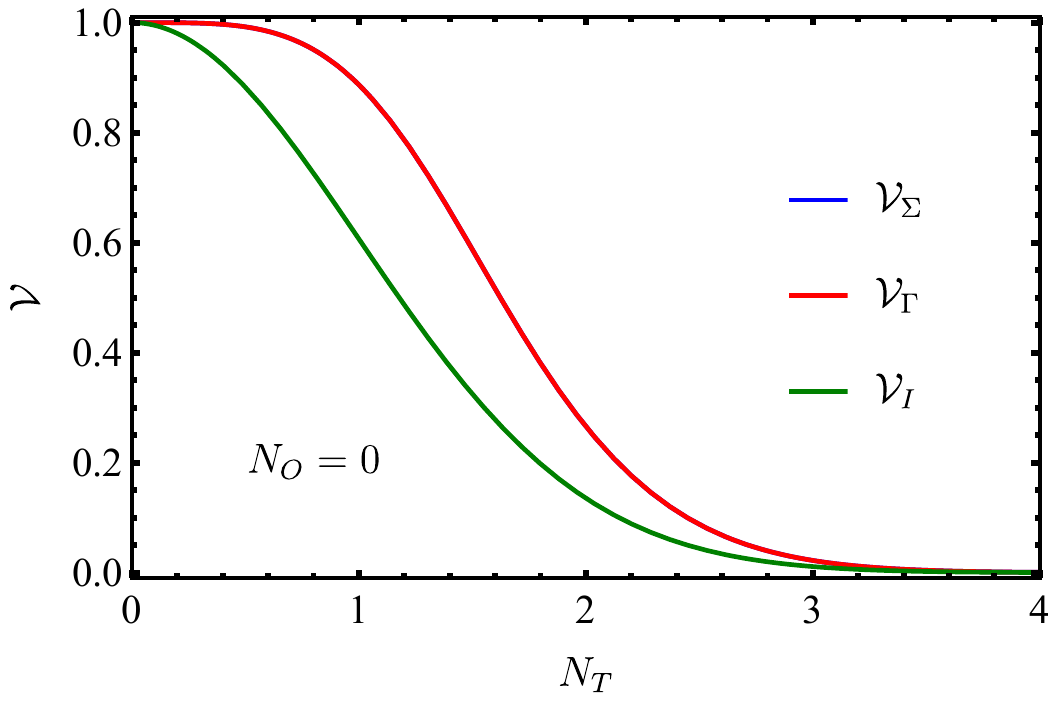}
     \caption{$N_O=0$}
    \end{subfigure}
    \hfill
    \begin{subfigure}[h]{.49\textwidth}
     \centering 
 \includegraphics[width=\textwidth]{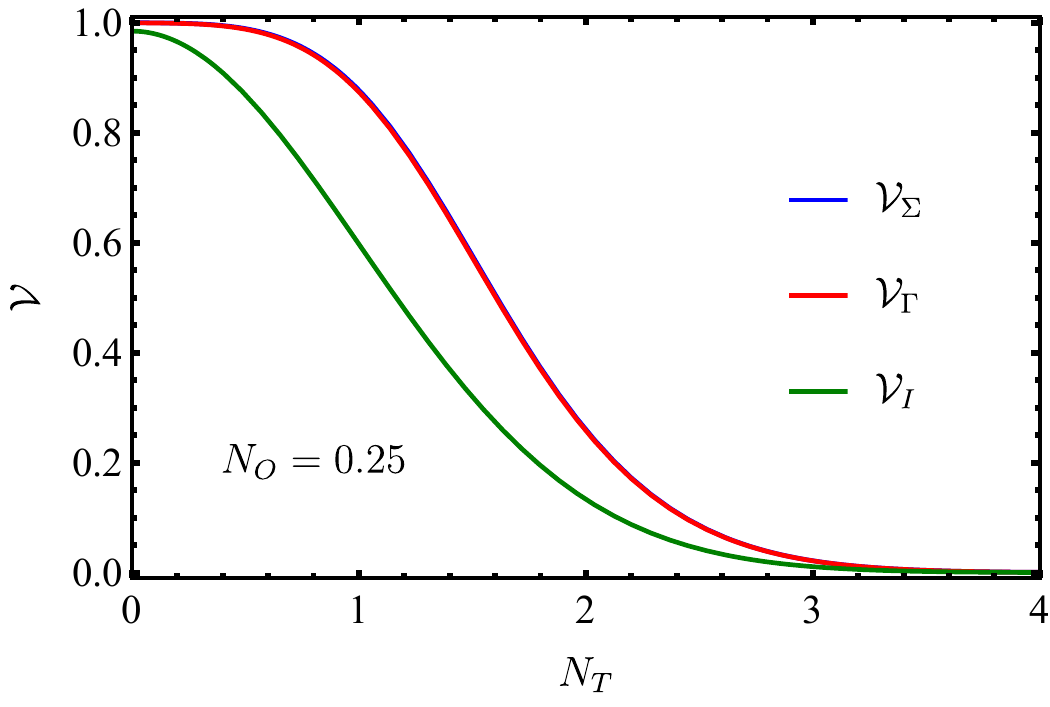}
     \caption{$N_O=0.25$}
    \end{subfigure}
    \begin{subfigure}[h]{.49\textwidth}
     \centering 
     \includegraphics[width=\textwidth]{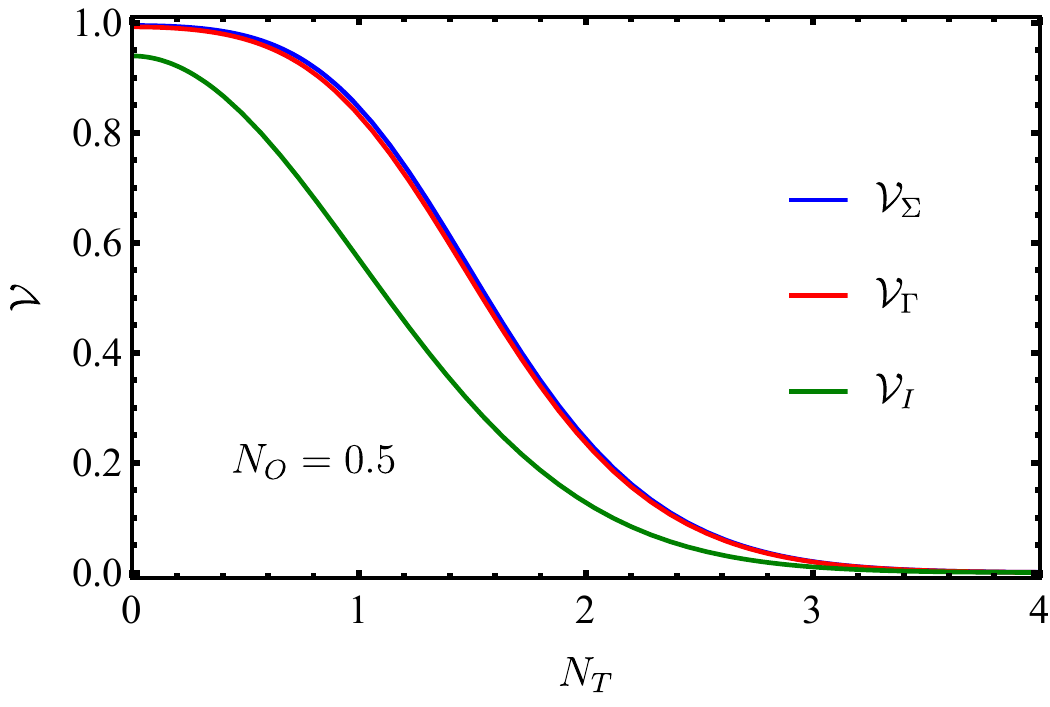}
     \caption{$N_O=0.5$}
    \end{subfigure}
    \hfill
    \begin{subfigure}[h]{.49\textwidth}
     \centering 
 \includegraphics[width=\textwidth]{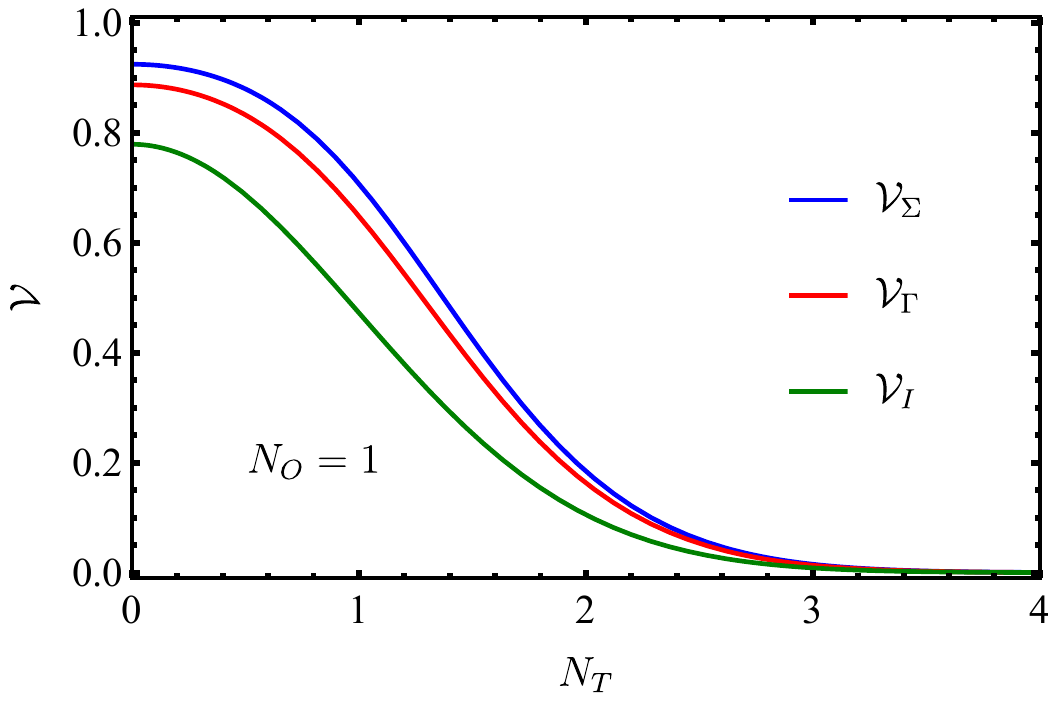}
     \caption{$N_O=1$}
    \end{subfigure}
    \caption{Visibilities of the integrated correlation image ($\mathcal{V}_\Sigma$), the pointwise correlation image, captured at $\br_b=0$ ($\mathcal{V}_\Gamma$), and the direct intensity image ($\mathcal{V}_I$) of a periodic intensity pattern $I_S(\br_s) \propto \cos^2 ( x_s / 2 w_s)$. The visibility is reported as a function of the dimensionless parameter $N_T=f_O/(k w_s \sigma_T)$, that quantifies the impact on strong turbulence on imaging. The four panels correspond to different values of $N_O=f_O/(k w_s \sigma_O)$, quantifying the number of resolution cells determined by the objective lens on a length $w_s$.}
    \label{fig:visibility}
\end{figure}

A suitable figure of merit to determine and compare the properties of each of the three images \eqref{eq:GammaRes}-\eqref{eq:SigmaRes}-\eqref{eq:IRes}, which we generally denote as $\mathfrak{F}(x)$ considering only the dependence on the relevant coordinate, is the visibility
\begin{equation}\label{eq:vis}
    \mathcal{V}_{\mathfrak{F}} =
    \frac{\mathfrak{F}(0)-\mathfrak{F}(-\pi w_s M_T)}{\mathfrak{F}(0)+\mathfrak{F}(-\pi w_s M_T)}, 
\end{equation}
where the reference image points correspond to the peak in $x_s=0$ and to the first minimum in $x_s=\pi w_s$ of the sample profile. Considering for definiteness $\Gamma(\br_a,0)$, namely the image corresponding to the peak value on $\mathrm{D}_b$, one obtains
\begin{align}\label{eq:viss}
    \mathcal{V}_{\Gamma} & =\frac{2\mathrm{e}^{-\frac{N_{T}^{2}+N_{O}^{2}}{2}}\cosh\frac{N_{O}x_{b}}{M_{L}\sigma_{O}}}{1+\mathrm{e}^{-\left(N_{T}^{2}+N_{O}^{2}\right)}\cosh^{2}\frac{N_{O}x_{b}}{M_{L}\sigma_{O}}}\Biggl|_{x_{b}=0}=\frac{2\mathrm{e}^{-\frac{N_{T}^{2}+N_{O}^{2}}{2}}}{1+\mathrm{e}^{-\left(N_{T}^{2}+N_{O}^{2}\right)}},
    \\
    \mathcal{V}_{\Sigma} & =\frac{4\mathrm{e}^{-\frac{3N_{O}^{2}+4N_{T}^{2}}{8}}}{2+\mathrm{e}^{-N_{T}^{2}-N_{O}^{2}}\left(1+\mathrm{e}^{\frac{N_{O}^{2}}{2}}\right)},
    \\
    \mathcal{V}_{I} & =\mathrm{e}^{-\frac{N_{O}^{2}+2 N_{T}^{2}}{4}}.
\end{align}
A comparison of the three visibilities is reported in Figure~\ref{fig:visibility}, as a function of $N_T$, for different fixed values of $N_O$. These plots can be interpreted as showing the detrimental change in image quality at fixed object and lens features, due to a decreasing turbulence correlation length (inversely proportional to $N_T$). Let us first consider for clarity the limit $N_O\to 0$, in which the finite lens aperture has no practical effect on the object resolution. In this case,
\begin{equation}
    \mathcal{V}_{\Sigma} = \mathcal{V}_{\Gamma} ,
\end{equation}
since integrating the objective image on $\mathrm{D}_b$ becomes equivalent to add images with identical properties, with the only effect of increasing the signal-to-noise ratio \cite{scala2019signal,massaro2022comparative}. On the other hand, considering
\begin{equation}
    \frac{\mathcal{V}_{\Gamma}}{\mathcal{V}_I} = \frac{2}{1 + \exp(-N_T^2)} ,
\end{equation}
it is evident that correlations provide better resolved images in any case in which turbulence is present, with the performance gap improving with increasing $N_T$. As $N_O$ increases, the both correlation images still outperform the direct intensity image. At the same time, a slight difference in favor of the integrated correlation image $\Sigma$ appears, due to the fact that, when the finite lens aperture becomes relevant, the contribution of rays passing from peripheral parts of the objective provide an increase in the average visibility.

\begin{figure}
    \centering
    \includegraphics[width=0.8\linewidth]{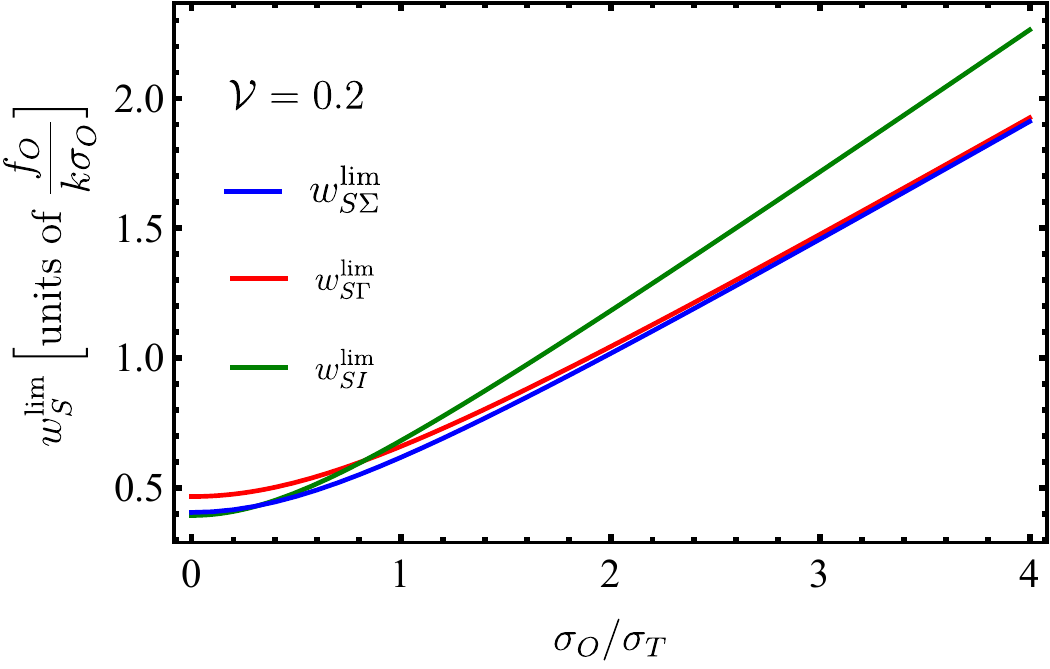}
    \caption{Limiting value of $w_s$, proportional to the spatial period of a sample intensity patterns, above which the visibility of the image pattern is larger than $20\%$, evaluated for integrated correlation imaging ($w_{s\Sigma}^{\lim}$), pointwise correlation imaging with $\br_b=0$ ($w_{s\Gamma}^{\lim}$) and intensity imaging (($w_{sI}^{\lim}$). Higher values correspond to smaller resolutions of the considered imaging methods. For any fixed $\sigma_O/\sigma_T\gtrsim 1$, correlation imaging (either pointwise and integrated) allows to resolve closer object peaks than direct intensity imaging.}.
    \label{fig:imagebroaden}
\end{figure}

A complementary analysis consists in fixing the lens ($\sigma_O$) and turbulence ($\sigma_T$) properties, and searching for the limiting value $w_S^{\lim}$ above which the pattern image is resolved with visibility larger than a given $\mathcal{V}$. Such a limit can be analytically obtained for the pointwise correlation image
\begin{equation}\label{eq:wlimGamma}
    w_{s\Gamma}^{\lim} = \frac{f_O}{k\sigma_O} \sqrt{ \frac{1 + (\sigma_O/\sigma_T)^2 }{ 2  \left| \log \frac{ 1 - \sqrt{1 - \mathcal{V}^2 } }{\mathcal{V}} \right| }  }
\end{equation}
and for the intensity image
\begin{equation}\label{eq:wlimI}
    w_{sI}^{\lim} = \frac{f_O}{2 k\sigma_O} \sqrt{ \frac{ 1 + 2 (\sigma_O/\sigma_T)^2 }{ \left| \log \mathcal{V} \right| } } ,
\end{equation}
allowing for an immediate comparison,
\begin{equation}\label{eq:wratio}
    \frac{w_{s\Gamma}^{\lim}}{w_{sI}^{\lim}} = \sqrt{ \frac{2 + 2 (\sigma_O/\sigma_T)^2 }{1 + 2 (\sigma_O/\sigma_T)^2}  \left| \frac{ \log \mathcal{V} }{ \log \frac{ 1 - \sqrt{1 - \mathcal{V}^2 } }{\mathcal{V}} }  \right| } \stackrel{ \sigma_O \gg \sigma_T } \longrightarrow   \left| \frac{ \log \mathcal{V} }{ \log \frac{ 1 - \sqrt{1 - \mathcal{V}^2 } }{\mathcal{V}} }  \right|^{\frac{1}{2}}
\end{equation}
Notice that the dimensional factor $f_O/k\sigma_O$ appearing in both \eqref{eq:wlimGamma} and \eqref{eq:wlimI} corresponds to the length scale of the resolution cell set by the objective. In the limit $\sigma_O\gg \sigma_T$, where the impact of turbulence on resolution is much more relevant than the lens size, the ratio in Eq.~\eqref{eq:wratio} is always smaller than one, implying that, whatever the threshold value $\mathcal{V}$ chosen for visibility, the pointwise correlation image is able to resolve finer details. The same reasoning applies for the integrated correlation image, whose limit $w_{s\Sigma}^{\lim}$ cannot be generally obtained analytically, since its quality coincides with that of $\Gamma$ in the considered $\sigma_O\gg \sigma_T$ regime. A numerical evaluation of the resolution limits at $\mathcal{V}=20\%$ for the three image functions is reported in Figure~\ref{fig:imagebroaden}. While at large $\sigma_O/\sigma_T$ the plot matches the expectations coming from the analytical results, at $\sigma_O/\sigma_T\ll 1$, when the turbulence tends to become irrelevant, the pointwise correlation is disadvantaged with respect to the other functions, due to well-known coherence effects that are mitigated by integration on the objecive lens \cite{scattarella2022resolution,massaro2022effect,scattarella2023periodic} In the case of $\sigma_O/\sigma_T\gtrsim 1$, it is possible to observe that both correlation imaging methods outperform direct intensity imaging in terms of resolution.

\section{Conclusions and outlook}\label{sec:conclusions}

The comparison between direct and correlation imaging in the presence of strong turbulence unambiguously revealed an advantage of the latter, either in the pointwise or in the integrated versions. This advantage, quantified by analyzing the visibility of periodic sample patterns, is more striking as the presence of turbulence becomes dominant in determining the image resolution. Though we considered a specific embodiment of CPI, that is an intrisically three-dimensional imaging techniques, in this work we limited ourselves to the case of an object in a specific plane, such that the image of the considered sample is focused. This choice allowed to decouple the comparison between robustness against turbulence of direct and correlation imaging from the overwhelming advantage of CPI in terms of resolution of out-of-focus objects. Nonetheless, future research will be devoted to characterize the effect of propagation in turbulent media on volumetric resolution in CPI devices.

While the present work characterized the differences between imaging techniques in terms of resolution, thereby highlighting the advantages of correlation imaging, other figures of merit are worth investigating, especially concerning signal-to-noise ratio. While the latter is typically a weak point of correlation imaging, recent theoretical results \cite{scala2019signal,massaro2022comparative} and experimental findings \cite{massaro2022lightfield,massaro2023correlated} showed that correlation imaging can compete with direct imaging in setups where at least one of the detector collects the focused image of a plane in the scene.
On the other hand, we will extend the analysis performed in this work, where we considered a strong-turbulence model, to different turbulence regimes, exploring diversified application ranges, especially in remote sensing and biological imaging.

\bmhead{Acknowledgements}
G.M, M.D, and F.V.P. acknowledge funding from Universit\`a degli Studi di Bari under project
ADEQUADE. Project ADEQUADE has received funding from the European Defence Fund (EDF)
under grant agreement EDF-2021-DIS-RDIS-ADEQUADE (n.\ 101103417). G.S. and C.L. acknowledge funding from the European Union's Horizon Europe research and innovation program under the project ``Quantum Secure Networks Partnership'' (QSNP, grant agreement No.\ 101114043). All authors acknowledge funding from INFN through the projects QUANTUM and QUISS. \\Funded by the European Union. The views and opinions expressed are, however, those of the authors only and do not necessarily reflect those of the European Union or the European Commission. Neither the European Union nor the granting authority can be held responsible for them.


\end{document}